\documentclass[aps,prfluids,superscriptaddress,longbibliography]{revtex4-2}

\usepackage{graphicx}
\usepackage{dcolumn}
\usepackage{bm}
\usepackage{color}
\usepackage[colorlinks=true,allcolors=blue,breaklinks=true]{hyperref}
\usepackage{amsmath} 

\begin{document}

\title{Effect of sub-critical fluid shear flow on granular bed strength}

\author{Dong Wang}
\affiliation{Department of Mechanical Engineering, Yale University, New Haven, Connecticut 06520, USA}

\author{Sophie Bodek}
\affiliation{Department of Civil and Environmental Engineering, Stanford University, Stanford, CA 94305, USA}

\author{Nicholas T. Ouellette}
\affiliation{Department of Civil and Environmental Engineering, Stanford University, Stanford, CA 94305, USA}

\author{Mark D. Shattuck}
\affiliation{Benjamin Levich Institute and Physics Department, The City College of New York, New York, New York 10031, USA}

\author{Corey S. O'Hern}
\email{corey.ohern@yale.edu}
\affiliation{Department of Mechanical Engineering, Yale University, New Haven, Connecticut 06520, USA}
\affiliation{Department of Physics, Yale University, New Haven, Connecticut 06520, USA}
\affiliation{Department of Applied Physics, Yale University, New Haven, Connecticut 06520, USA}
\affiliation{Department of Materials Science, Yale University, New Haven, Connecticut 06520, USA}

\date{\today}

\begin{abstract}
Interactions between fluids and granular materials are prevalent on the Earth's surface. In the case of fluid flow over a sediment bed, the fluid imparts a shear stress to the granular materials. When the applied shear stress is above a critical value, the grains become entrained in the fluid flow. Prior experimental studies have shown that granular beds subjected to a sub-critical fluid flow can strengthen in the same direction as the sub-critical flow. In contrast, granular beds can become weaker in the direction opposite to the sub-critical fluid flow. To investigate the grain-scale mechanisms that control directional strengthening and weakening, we perform discrete element method (DEM) simulations of granular beds subjected to model fluid flows in two (2D) and three (3D) dimensions with varied inter-particle static friction coefficients and conditioning flow speeds. In these studies, the sub-critical grain motion does not cause significant bed compaction. Instead, we find that the strength of a granular bed in a particular direction is highly correlated with the fraction of {\it surface} grains that can be dislodged by a fluid force applied in that direction. Further, the anisotropic bed strength only persists over a finite time scale that is set by the Shields number.  We also show that inter-particle static friction is not required for bed strength anisotropy, but varying the friction affects the magnitude of the anisotropy. This research enhances the grain-scale understanding of erosion of granular beds caused by fluid flows and underscores the importance of tracking the history of the fabric of the bed surface since it couples strongly to bed strength.
\end{abstract}

\maketitle

\section{Introduction}
\label{sec:intro}

Much of the Earth's surface is composed of granular materials~\cite{jerolmack19natrevphys} that are continually exposed to hydrodynamic and aerodynamic driving forces from water and wind. When these flows are sufficiently strong, they can dislodge individual grains, followed by groups of grains, and subsequent erosion of the bed~\cite{galay83wrr, buffington97wrr, pahtz20revgeophys, dey14}. Erosion can modify the landscape of the Earth's surface, such as re-shaping riverbeds, shorelines, and dunes~\cite{poesen18espl, wu17geomorphology, vellinga82coastaleng, feagin05frontecoenv}, mobilize organic carbon from terrestrial vegetation and deposit it into rivers and sediment beds~\cite{berhe18areps, hilton20natrevearthenv}, and destroy and then form new habitats for fish, aquatic microbes, and vegetation~\cite{im11ecoeng}. Therefore, understanding and predicting erosion is crucial to many physical, geological, agricultural, and ecological processes. 

Predicting the onset of erosion in fluid-driven granular beds is difficult for several reasons. First, erosion involves the coupling of fluid dynamics, which is often turbulent, and granular mechanics, which involves spatially heterogeneous force propagation.  In addition, the onset of erosion depends on the history of how granular beds are formed~\cite{buffington97wrr, paphitis2005sedimentology, haynes2007jhydrauleng, piedra2012sedimentology, mao2018geomorphology, masteller19grl, pahtz20revgeophys, masteller2025esd}. Several mechanisms have been proposed to explain the history dependence of erosion onset. For example, the critical stress necessary to dislodge fine grains is less than that for coarse grains in polydisperse granular beds. Therefore, stresses slightly below the critical value for the bulk system can first entrain finer grains into the fluid flow and remove them from the top layers of the bed, leaving only coarser grains with larger values for the critical stress. As a result, the bed can ``armor''~\cite{dietrich1989sediment, ferdowsi17natcomm, allen2018prf}, i.e. become stronger as small grains are removed from the top layers of the bed. Even without entraining grains in the fluid flow, weak fluid driving can give rise to velocity fluctuations in the grains, which can lead to compaction and strengthening of the bed~\cite{paphitis2005sedimentology, charru2004jfm, allen2018prf}.

\begin{figure}[t!]
    \centering
    \includegraphics[width=\linewidth]{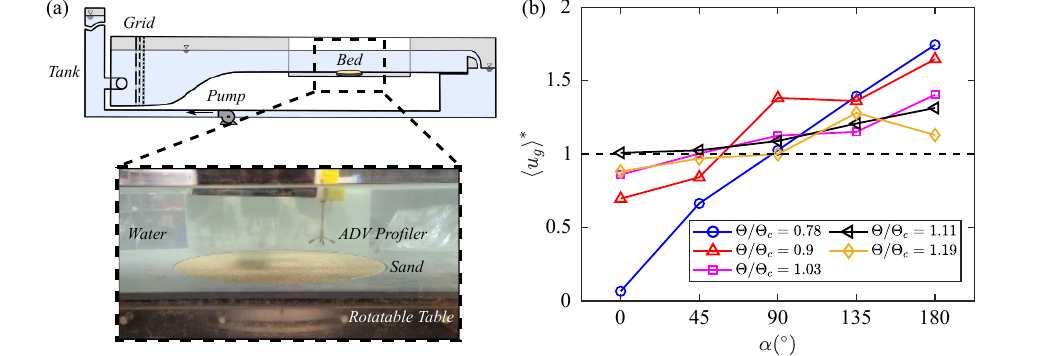}
    \caption{(a) Top panel: Sketch of the experimental setup of fluid-driven sediment beds. The blue-shaded regions contain water. The dashed box highlights the location of the bed of sand. The water is driven by a pump through the tank and then the grid to produce a desired shear stress at the surface of the granular bed. Bottom panel: Close-up of the region highlighted by the dashed box in (a), which includes the sand bed, rotatable table, acoustic Doppler velocimeter (ADV) profiler, and water. (b) Mean grain speed, normalized by that measured in the control condition, $\left\langle u_g \right\rangle^*$, plotted as a function of the angle between the conditioning flow and the subsequent flow to determine bed strength, $\alpha$, for several normalized Shields numbers $\Theta/\Theta_c$ near the critical value. The horizontal dashed line corresponds to $\left\langle u_g \right\rangle^* = 1$.}
    \label{fig:intro}
\end{figure}

Both armoring and compaction do not induce strong anisotropy in granular bed strength~\cite{majmudar05nature, behringer18rpp, wang20natcomm}. However, in recent experimental studies, we demonstrated that granular bed strength can depend on the direction in which the fluid stresses are applied~\cite{Galanis2022prf, bodek2026jgres}. For example, in Fig.~\ref{fig:intro} (a), we illustrate a granular bed of natural sand subjected to sub-critical conditioning fluid flows. The mean surface grain speed normalized by that measured in the control bed, $\langle u_g \rangle^*$, varies with the angle $\alpha$ between the conditioning flow and the subsequent fluid flow used to determine the bed strength. In particular, as shown in Fig.~\ref{fig:intro} (b), $\langle u_g \rangle^*$ is lower at $\alpha =0^{\circ}$ and higher at $180^{\circ}$, which indicates that the conditioned bed becomes stronger (weaker) in the same (opposite) direction as the conditioning fluid flow~\cite{bodek2026jgres}. The magnitude of the anisotropy decreases as the fluid driving approaches and exceeds the critical Shields number. We have found that when the beds are weaker, they possess an elevated fraction of highly mobile grains. However, these prior studies were not able to identify signatures in the interparticle contact networks of the bed that give rise to the increased fraction of mobile grains, which would enable predictions of anisotropic bed strength. 

In this work, we will elucidate the structural changes in the granular beds that are caused by the conditioning fluid flows and give rise to aniostropic bed strength via discrete element method (DEM) simulations of packings of frictional spherical grains subjected to model fluid shear flows. We first show that the DEM simulations recapitulate the anisotropic strength of granular beds. In particular, the mean speed of the grains at the bed surface for the conditioned bed is lower (higher) in the same (opposite) direction as the conditioning flow. We then show that differences in the mean grain speed arising from different flow directions vanish in the long-time limit and as the fluid driving approaches and exceeds the critical Shields number, which emphasizes the cases where the fluid flow can erase, rather than encode the flow history of the bed. We then show that the mean grain speed is strongly correlated with the fraction of grains with small values for the minimum force $F_{g,c}$ required to dislodge a single grain while fixing all other grains in the bed. Further, we find that inter-grain static friction promotes bed strength anisotropy, but it is not required for nonzero bed strength anisotropy. Finally, we demonstrate that the angular dependence of $F_{g,c}$ is largely determined by the inter-grain fabric, not the inter-grain force network, among surface grains.

The remainder of the article is organized as follows. In Sec.~\ref{sec:methods}, we describe the DEM simulations of granular beds driven by a coarse-grained fluid model in both two (2D) and three (3D) dimensions, how we prepare granular beds with and without sub-critical conditioning fluid flows, and the analysis methods used to characterize granular bed strength. In Sec.~\ref{sec:results}, we present the results from the DEM simulations on how the mean speed and $F_{g,c}$ of the surface grains change as a function of the conditioning flow speed, how inter-grain friction tunes the magnitude of the bed strength anisotropy, and how the angular dependence of $F_{g,c}$ for individual grains can be calculated analytically from the inter-grain fabric. In Sec.~\ref{sec:summary}, we summarize the conclusions and propose promising directions for future research, such as designing protocols to create granular beds with maximum strength anisotropy. We also include five Appendices that describe the calculation of the critical Shields number to erode granular beds, the choice of the time window used for averaging the grain speed, the magnitude of compaction of granular beds during the conditioning flows, differences in the structural properties of sedimented beds generated using varied protocols, and the derivation of an analytical expression for the minimum force required to dislodge a surface grain from the granular bed.

\section{Methods}
\label{sec:methods}

\begin{figure}[t!]
    \centering
    \includegraphics[width=\linewidth]{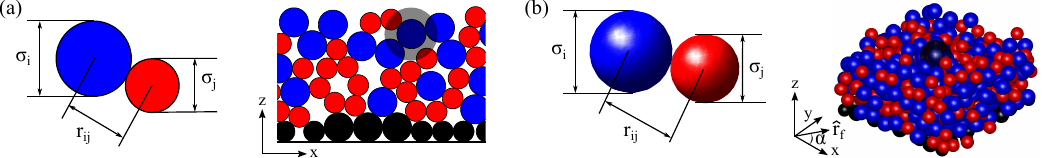}
    \caption{Schematic diagrams of the DEM simulations of fluid-driven granular beds in (a) 2D and (b) 3D. In (a) and (b), the left panels provide details of the grain interactions: $\sigma_i$ and $\sigma_j$ are the diameters of grains $i$ and $j$, and $r_{ij}$ is the separation between the centers of grains $i$ and $j$. The right panels give examples of settled granular beds with associated Cartesian coordinate axes. A gray circle in (a) and sphere in (b) indicate example regions used to calculate the local packing fraction at the bed surface. In the right panels of (a) and (b), the black grains indicate grains in the basal layer with fixed positions and orientations, and the blue and red grains are mobile large and small grains, respectively. In the right panel of (b), $\hat{r}_f$ indicates the direction of the fluid flow and $\alpha$ is the angle between $\hat{r}_f$ and the positive $x$-axis.}
    \label{fig:model}
\end{figure}

We employ discrete element method (DEM) simulations of granular beds (i.e. static packings of spherical particles) subjected to fluid shear flows using a coarse-grained model for the fluid in 2D and 3D~\cite{clark2015onset, clark2017prfluids}. The granular bed consists $N$ grains modeled as soft, frictional disks (spheres) in 2D (3D). We consider mixtures of half large and half small grains with a diameter ratio of $\sigma_l/\sigma_s = 1.4$ in 2D and $1.2$ in 3D. These diameter ratios have been selected to inhibit crystallization and maintain disordered granular beds during preparation and applied fluid flow in 2D and 3D~\cite{ohern2003pre, speedy1999jcp, zhang2014pre_bmg}. We set -${\hat z}$ as the direction of gravity, and employ periodic boundary conditions in the $x$-direction (both $x$- and $y$-directions with a square shape) in 2D (3D). The bed is confined by a fixed bottom wall at $z = 0$ and has an open boundary at the top surface. We use the Cundall-Strack (CS) model to implement inter-grain friction with static friction coefficient $\mu$, where grains have a geometrically smooth surface and friction arises from spring forces that are proportional to the relative tangential displacements of overlapping grains~\cite{cundall-strack}. Schematics of the simulation models in 2D and 3D are shown in Fig.~\ref{fig:model}. For the CS friction model, the pairwise normal force on grain $i$ from $j$ due to overlapping grains $i$ and $j$ is 
\begin{equation}
    \vec{F}_{ij}^n = K (\sigma_{ij} - r_{ij}) \theta(\sigma_{ij} - r_{ij}) \hat{r}_{ij},
\end{equation}
where $K$ is the spring constant, $r_{ij}$ is the separation between the centers of grains $i$ and $j$, $\sigma_{ij}$ is the sum of the radii of grains $i$ and $j$, and $\hat{r}_{ij}$ is the unit vector that points from grain $j$ to $i$, and $\theta(\cdot)$ is the Heaviside step function that ensures that only repulsive forces are nonzero for $r_{ij} < \sigma_{ij}$ and $F^n_{ij}= 0$ for $r_{ij} \ge \sigma_{ij}$. For the CS model, the tangential force on grain $i$ arising from $j$ is 
\begin{equation}
        \vec{F}_{ij}^t = \min(\mu F_{ij}^n, K_t \zeta_{ij})\hat{t}_{ij},
\end{equation}
where $\min(A,B)$ selects the minimum of $A$ and $B$, $K_t = K/3$ is the spring constant of the tangential spring, $\zeta_{ij}$ is the spring elongation in the tangential direction that occurs during the contact between grains $i$ and $j$, and $\hat{t}_{ij}$ is the direction of the tangential displacement with ${\hat r}_{ij} \cdot {\hat t}_{ij} = 0$. $\zeta_{ij}$ is obtained by integrating 
\begin{equation}
    \frac{d\vec{\zeta}_{ij}}{dt} = \vec{v}_i - \vec{v}_j - \left( (\vec{v}_i - \vec{v}_j) \cdot \hat{r}_{ij} \right) \hat{r}_{ij} + \frac{1}{2} (\vec{\omega}_i + \vec{\omega}_j) \times \vec{r}_{ij} - \frac{\vec{\zeta}_{ij} \cdot (\vec{v}_i - \vec{v}_j)}{r_{ij}} \hat{r}_{ij}
\end{equation}
over the lifetime of the contact between grains $i$ and $j$, where $\vec{v}_i$ and $\vec{v}_j$ are the translational velocities of grains $i$ and $j$, $\vec{r}_{ij} =r_{ij} {\hat r}_{ij}$, and $\vec{\omega}_i$ and $\vec{\omega}_j$ are angular velocities of grains $i$ and $j$, respectively. We also include a dissipative force on grain $i$ from colliding with grain $j$:
\begin{equation}
    \vec{F}_{ij}^d = \gamma_v \frac{m_i m_j}{m_i + m_j} \left( (\vec{v}_i - \vec{v}_j) \cdot \hat{r}_{ij} \right) \hat{r}_{ij},
\end{equation}
where $\gamma_v = -2\ln (e_n / t_c)$ is the dissipation rate, $e_n$ is the coefficient of restitution, $t_c = \pi \sqrt{m_i m_j / (m_i + m_j) /K}$ is the grain-grain collision time scale, and $m_i$ and $m_j$ are the masses of grains $i$ and $j$.

We use a coarse-grained fluid model that can accurately predict the critical Shields number to erode granular beds as a function of the shear Reynolds number~\cite{clark2017prfluids}. In this fluid model, we only consider drag forces exerted on each grain $i$ from the fluid flow: 
\begin{equation}
    \vec{F}_{f, i} = B_1 \left(v_{f,0} f(\phi_i) \hat{r}_f - \vec{v}_i \right) + B_2 | v_{f,0} f(\phi_i) \hat{r}_f - \vec{v}_i | \left(v_{f,0} f(\phi_i) \hat{r}_f - \vec{v}_i \right),
    \label{eq:fluid_force}
\end{equation}
where $v_{f,0}$ is the fluid speed at the bed surface, $\hat{r}_f$ is the direction of the applied fluid flow, $f(\phi_i)$ is the coarse-grained fluid velocity profile that depends on the local packing fraction $\phi_i$ near grain $i$, $B_1 = 3 \pi \rho_f \nu \sigma_i$ is the linear drag coefficient, $\rho_f$ is the mass density of the fluid, $\nu$ is the kinematic viscosity of the fluid, and $B_2$ is the quadratic drag coefficient with $B_2 = 0$ in 2D and $B_2 = \pi \rho_f \sigma_i^2 / 20$ in 3D. We only consider flows where $\hat{r}_f$ is perpendicular to the direction of gravity. In 3D, we denote the angle between $\hat{r}_f$ and the positive $x$-axis as $\alpha$, as illustrated in Fig.~\ref{fig:model} (b). To calculate $\phi_i$, we draw a test circle (sphere) centered at grain $i$ with diameter $\sigma_i + \sigma_l$ and determine $\phi_i$ to be the ratio of the total area (volume) of the enclosed grains divided by the area (volume) of the test circle (sphere) in 2D (3D). An example test circle and sphere to calculate $\phi_i$ are shown in Fig.~\ref{fig:model} (a) and (b). In 2D, $f(\phi_i) = e^{-b(\phi_i - \phi_m^{2D})}$ and in 3D, $f(\phi_i) = (e^{-b\phi_i} - e^{-b\phi_m^{3D}})/(e^{-b\phi_t^{3D}} - e^{-b\phi_m^{3D}})$. We set $b = 9$, $\phi_m^{2D} = 0.5$, $\phi_t^{3D} = 0.38$, and $\phi_m^{3D} = 0.5$. $\phi_m^{2D}$ ($\phi_m^{3D}$) corresponds to the maximum packing fraction in the bulk of the bed in 2D (3D), and $\phi_t^{3D}$ is the typical packing fraction at the surface of the bed in 3D. The functional form for $f(\phi_i)$ ensures that the velocity of the fluid decays to $0$ inside the bulk of the bed (with $b$ controlling the spatial decay rate) and mimics the linear increase of fluid velocity with the distance to the surface of the bed in the viscous layer. The fluid flow imposes shear stresses on the grains, and we characterize the shear stress at the surface of bed $\tau_f$ by calculating the Shields number,
\begin{equation}
    \Theta = \frac{\tau_f}{\rho_g g' \sigma_s} = \frac{2}{3}\frac{B_1v_{f,0} + B_2v_{f,0}^2}{m_s g'},
\end{equation}
where $m_s$ is the mass of the small grains, $\rho_g$ is the mass density of the grains, $g$ is the gravitational acceleration, and $g' = (\rho_g / \rho_f - 1) g$.

The positions, translational velocities, angular displacements and velocities of the grains in the granular bed are obtained by integrating the following two equations of motion:
\begin{equation}
    m_i \frac{d \vec{v}_i}{dt} = \vec{F}_{g, i} + \vec{F}_{f, i} - m_i g' \hat{z},
    \label{eq:eom_pos}
\end{equation}
\begin{equation}
    I_i \frac{d \vec{\omega}_i}{dt} = \vec{T}_{g, i},
    \label{eq:eom_rot}
\end{equation}
where $\vec{F}_{g, i} = \sum_{j \ne i=1}^{N} ( \vec{F}_{ij}^n + \vec{F}_{ij}^t + \vec{F}_{ij}^d)$ is the total force on grain $i$ from other grains, $I_i = m_i\sigma_i^2/8$ ($m_i\sigma_n^2/10$) is the moment of inertia of grain $i$ in 2D (3D), and $\vec{T}_{g, i} = \sum_{j \ne i=1}^{N} [\sigma_i/(2\sigma_{ij})] (\vec{F}_{ij}^t \times \vec{r}_{ij})$ is the total torque on grain $i$.

In both 2D and 3D, we study granular beds with $N = 800$ grains with an average bed height $H \approx 4\sigma_s$. We have shown that the critical Shields number $\Theta_c$ to erode the model granular beds does not depend strongly on $N$. For the grains, we set $e_n = 0.05$ and $g'/(\sigma_sK/m_s) = 10^{-4}$ to mimic rigid grains and vary the static friction coefficient $0 \le \mu \le 10$. For the fluid flow, we set $\rho_f/\rho_g = 2/3$ and viscosity $\nu / (\sigma_s^2 / \sqrt{m_s/K}) = 0.1$, and vary $v_{f,0}$ to tune the Shields number $0.1 < \Theta < 1.8$ and the particle Reynolds number $0.0002 < {\rm Re}_p < 0.003$, where ${\rm Re}_p = v_{f,0}\sigma_s / \nu$. We also note that $\Theta_c$ does not depend sensitively on $e_n$ for the ${\rm Re}_p$ values we study. We carry out flow studies for $10$ randomly generated granular beds for each $\mu$ and report the ensemble-averaged results for the average grain speed and other quantities. Below, we use $\sigma_s$ as the unit of length, $K\sigma_s$ as the unit of force, $K\sigma_s^2$ as the unit of torque, and $\sqrt{m_s/K}$ as the unit of time.

We start the DEM simulations by generating a static granular bed. We first place the grains on a square (cubic) lattice with lattice constant $l_c / \sigma_l = 1.01$ in 2D (3D). We then apply sinusoidal vibrations to the bottom plate in the $z$-direction: $z_b(t) = z_0\sin(\omega_bt)$ with $z_0/\sigma_s = 1$ and $\omega_b/(\sqrt{K/m_s}) = 1$ for a total time of $T_b / (\sqrt{m_s/K}) = 1000$ at $v_{f,0} = 0$ and $\nu / (\sigma_s^2 / \sqrt{m_s/K}) = 0$. We then stop the vibrations and allow the grains to settle at $v_{f,0} = 0$ and $\nu / (\sigma_s^2 / \sqrt{m_s/K}) = 0.1$. We deem that the bed has settled when the following four stopping criteria are met: (1) the net force on each grain is smaller than $10^{-10}$, (2) the net torque on each grain is smaller than $10^{-10}$, (3) the translational velocity of each grain is smaller than $10^{-10}$, and (4) the angular velocity of each grain is smaller than $10^{-10}$.  We then fix the positions and orientations of grains that are in contact with the bottom plate to mimic a rough bottom. This protocol to generate granular beds ensures that the local packing fraction profile does not depend strongly on $\mu$~\cite{an2008powdertech, gago2016paperphys} and that compaction by the applied fluid flows (as determined by average changes in the packing fraction of the surface grains) is small with $\Delta \langle \phi \rangle \lesssim 10^{-2}$. 

After we create the settled granular beds, we determine $\Theta_c$ for each bed by applying the fluid flow over a range of $\Theta$ at $\alpha = 0^{\circ}$ for a time $T_d = 10^7$. We choose $T_d = 10^7$ so that if grains were entrained in the fluid flow near $\Theta \sim \Theta_c$, they would move more than $10$ times the simulation domain in the flow direction, enabling the system to achieve a steady state. We also track grains at the surface of the bed. To identify the surface grains, we introduce a probe grain with diameter $0.5\sigma_s$, at a given $x$ or $(x, y)$ coordinate in 2D (3D), and height $z = H + 10\sigma_s$, and slowly reduce its $z$-position until it contacts one of the grains in the bed, which is then deemed to be a surface grain. We break the horizontal domain evenly into bins with size $0.25\sigma_s$ (area $0.25^2 \sigma_s^2$) in 2D (3D) and repeat this process by initializing a probe grain in each bin.  This procedure identifies $N_{\mathrm{surf}}$ surface grains and $N_{\mathrm{bulk}} = N - N_{\mathrm{surf}}$ grains in the bulk. We assume that $\Theta < \Theta_c$ when the following two criteria are satisfied: (1) the four stopping criteria above for the grains are satisfied and (2) less than $10\%$ of the surface grains have moved more than $\sigma_l$. To generate a conditioned granular bed, we apply a fluid flow with $\Theta_p \le \Theta_c$ to the original settled bed (i.e. the ``control'') for a time $T_d$, and allow grains to come to rest, such that all four stopping criteria for the grains are met. All measurements for the control beds are performed with $\alpha = 0^{\circ}$.

\begin{figure}[t!]
    \centering
    \includegraphics[width=\linewidth]{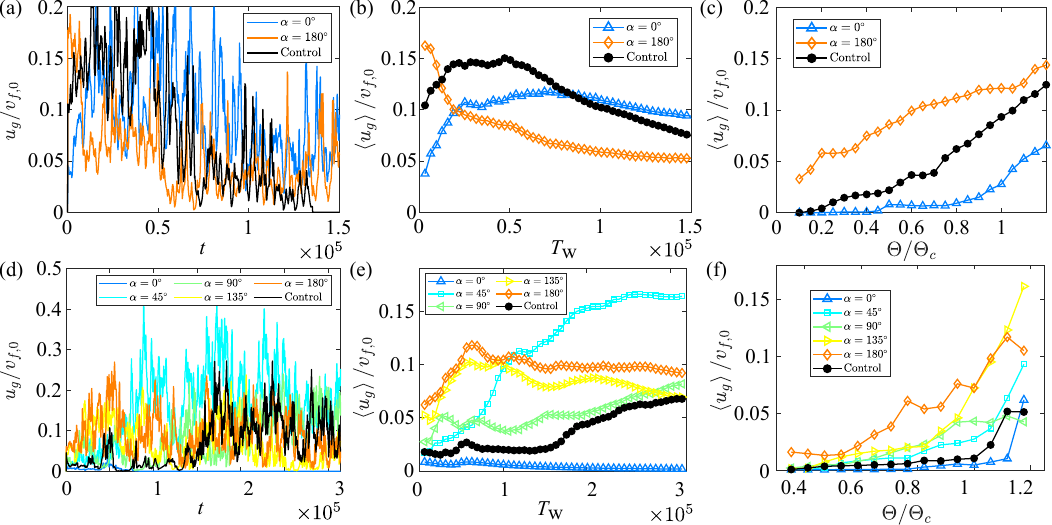}
    \caption{(a) Mean grain velocity normalized by the fluid speed at the bed surface, $u_g/v_{f,0}$, plotted versus time $t$ for the control and conditioned beds at $\alpha = 0^{\circ}$ and $180^{\circ}$ in 2D. The conditioning flow is set to $\Theta_p / \Theta_c = 0.8$. (b) Time-averaged mean grain speed normalized by the fluid speed at the bed surface, $\left\langle u_g \right\rangle /v_{f,0}$, plotted versus the averaging time window $T_w$ for the same data in (a). (c) $\left\langle u_g \right\rangle /v_{f,0}$ at $T_w = 10^4$ plotted versus the normalized Shields number of the applied fluid flow, $\Theta / \Theta_c$. (d) $u_g/v_{f,0}$ plotted versus time $t$, for the control and conditioned beds for several values of $\alpha$ in 3D. The conditioning flow is set to $\Theta_p / \Theta_c = 0.8$. (e) $\left\langle u_g \right\rangle /v_{f,0}$ plotted versus $T_w$ for the same data in (d). (f) $\left\langle u_g \right\rangle /v_{f,0}$ at $T_w = 5 \times 10^4$ plotted versus $\Theta / \Theta_c$.}
    \label{fig:ug}
\end{figure}

For fluid-driven granular beds at a given time $t$ following the initiation of the fluid flow, we calculate the mean grain velocity $u_g = \sum_{i=1}^{N_{\mathrm{surf}}} \vec{v}_i \cdot \hat{r}_f / N_{\mathrm{surf}}$ of grains at the surface of the bed to assess its strength. We can also calculate the time-averaged mean grain velocity $\left\langle u_g \right\rangle = \int_{0}^{T_w} u_g dt / T_w$, where $T_w$ is the averaging time window. For the control and conditioned granular beds, we also calculate the minimum force $F_{g,c}$ required to dislodge a single surface grain from the bed (i.e. move the grain by $\ge \sigma_l$) given that all other grains are fixed. We also calculate the local packing fraction for the surface grains and the characteristic Shields number $\Theta_{g,c}$ to dislodge the surface grain that determines $F_{g,c}$ using Eq.~\ref{eq:fluid_force}. After calculating $\Theta_{g,c}$, we can obtain the fraction of surface grains that will be dislodged by a fluid flow with $\Theta$, $M(\Theta) = N(\Theta > \Theta_{g,c}) / N_{\mathrm{surf}}$, where $N(\Theta > \Theta_{g,c})$ is the number of grains with $\Theta > \Theta_{g,c}$ and the characteristic fraction of mobile grains at erosion onset is $M_c = M(\Theta = \Theta_c)$.

\section{Results}
\label{sec:results}

In this section, we describe the main results of the DEM simulations of the anisotropic strength of fluid-driven granular beds. In Sec.~\ref{sec:meanspeed}, we show that the mean surface grain speed $\left\langle u_g \right\rangle$ depends on the direction of the conditioning fluid velocity for subcritical Shields number $1 \gtrsim \Theta_p / \Theta_c \gtrsim 0.6$, and that the angle dependence of $\left\langle u_g \right\rangle$ vanishes when the timescale for the applied fluid flow diverges $T_w \rightarrow \infty$. In Sec.~\ref{sec:singlegrain}, we show that $\langle u_g \rangle$ is controlled by the fraction of highly mobile surface grains, which can be determined by the minimum force $F_{g,c}$ to dislodge each surface grain individually when all others are held fixed. We then show that the conditioning fluid flows can alter $F_{g,c}$, which causes the angular dependence of the fraction of mobile grains and mean grain speed. In addition, we show that inter-grain static friction is not necessary to generate bed-strength anisotropy, but varying the static friction coefficient $\mu$ changes the magnitude of the anisotropy. In Sec.~\ref{sec:compaction}, we show that the change in the average packing fraction at the surface of the granular beds due to the conditioning fluid flows is negligible. Finally, in Sec.~\ref{sec:fabric}, we emphasize that $F_{g,c}$ for granular beds is determined mainly by the inter-grain fabric, not by the inter-grain forces.

\subsection{Mean Grain Speed Depends on the Direction of the Conditioning Fluid Flows}
\label{sec:meanspeed}

We investigate the strength of granular beds by calculating the mean surface grain speed $u_g$ when the bed is subjected to an applied fluid flow. Larger $u_g$ at a given $\Theta/\Theta_c$ reflects a weaker bed, whereas smaller $u_g$ at the same $\Theta/\Theta_c$ reflects a stronger bed. As shown in Fig.~\ref{fig:ug} (a), we find that $u_g$ depends on the direction of the applied conditioning fluid flow in 2D. Since the system sizes we study are relatively small, the mean surface grain speed fluctuates in time. To better compare the bed strength at different values of $\alpha$, we calculate the time-averaged mean grain speed $\left\langle u_g \right\rangle$. We find that $\left\langle u_g \right\rangle$ is larger (smaller) at $\alpha = 180^{\circ} (0^{\circ})$ for the conditioned beds compared to the control for a small averaging time window, e.g., $T_w \lesssim 10^4$ in 2D as shown in Fig.~\ref{fig:ug} (b). In the large-$T_w$ limit, $\left\langle u_g \right\rangle$ for the control, $\alpha=0$ and $180^{\circ}$ conditioned beds approach each other. To determine the averaging time window $T_w^*$ for calculating $\left\langle u_g \right\rangle$ that reflects the initial structure of the bed, we investigate how the grain dynamics evolve as a function of time in Appendix~\ref{appendix:timescale}. We have found that the fraction of highly mobile grains exhibits an initial increase and subsequent decrease for $\Theta_p / \Theta_c \lesssim 0.7$ over a short time scale $T_w^* \approx 10^{4}$. We set the time window for the time-averaged grain speed to be $T_w^*$. Using the time window $T_w^*$, we find that anisotropy in bed strength (especially weakening in 2D) also diminishes with increasing $\Theta/\Theta_c$ near and above the critical Shields number as shown in Fig.~\ref{fig:ug} (c). We find similar results for bed strength anisotropy in 3D in Fig.~\ref{fig:ug} (d)-(f). In 3D, $\left\langle u_g \right\rangle$ increases monotonically with $\alpha$ for $0^{\circ} \le \alpha \le 180^{\circ}$ for small $T_w$ as shown in Fig.~\ref{fig:ug} (e). This result emphasizes that 3D granular beds are stronger in the conditioning flow direction and weaker in the direction opposite to that of the conditioning flow, which is consistent with the experimental findings introduced in Sec.~\ref{sec:intro}~\cite{Galanis2022prf, bodek2026jgres}. In 3D, the bed strength anisotropy in $\left\langle u_g \right\rangle$ persists for $\Theta/\Theta_c \lesssim 1.2$ as shown in Fig.~\ref{fig:ug} (f). Below, we set $T_w^* = 10^4$ in 2D and $5 \times 10^4$ in 3D when reporting $\left\langle u_g \right\rangle$. 

\begin{figure}[t!]
    \centering
    \includegraphics[width=\linewidth]{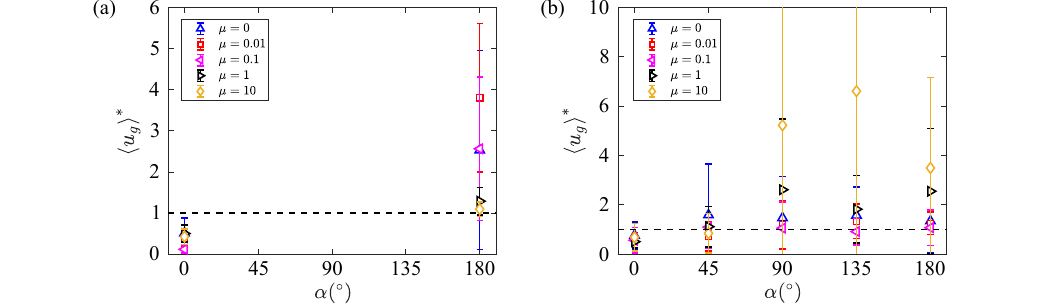}
    \caption{The ratio $\left\langle u_g \right\rangle^*$ of the time-averaged mean surface grain velocity for conditioned beds to that of the control bed at the same Shields number $\Theta$ plotted versus the flow direction $\alpha$ in (a) 2D and (b) 3D. In both (a) and (b), the conditioning fluid flow was $\Theta_p / \Theta_c = 0.8$ and the fluid flow for the control bed was at $\Theta / \Theta_c = 1$, $\left\langle u_g \right\rangle^*$ is calculated at $T_w = 10^4$ in 2D and $5 \times 10^4$ in 3D, we include a range of static friction coefficients $0 \le \mu \le 10$, and the dashed horizontal line indicates $\left\langle u_g \right\rangle^* = 1$. Data are calculated as an ensemble average over $10$ randomly generated configurations for each $\mu$.}
    \label{fig:ug_ratio_mu}
\end{figure}

As discussed above, bed strength anisotropy occurs for both 2D and 3D granular beds. In addition, bed strength anisotropy occurs for beds with both frictionless and frictional grains. In Fig.~\ref{fig:ug_ratio_mu}, we plot the time-averaged mean surface grain velocity $\left\langle u_g \right\rangle^*$ for the conditioned beds normalized by that for the control bed at $\Theta = \Theta_c$ over a range of static friction coefficients $0 \le \mu \le 10$ in 2D and 3D. $\langle u_g \rangle^*$ is smaller (larger) than $1$ at $\alpha = 0^{\circ}$ ($180^{\circ}$) in 2D and 3D. In addition, $\left\langle u_g \right\rangle^*$ increases non-monotonically from $\alpha = 0^{\circ}$ to $180^{\circ}$ in 3D, with a $\mu$-dependent maximum $\left\langle u_g \right\rangle^*$ at $90^{\circ} \le \alpha \le 180^{\circ}$. Interestingly, this non-monotonic behavior of $\left\langle u_g \right\rangle^*$ as a function of $\alpha$ has also been observed in experiments, e.g. $\left\langle u_g \right\rangle^*$ occurs at $\alpha = 135^{\circ}$ with $\Theta / \Theta_c = 1.19$ in Fig.~\ref{fig:intro} (b)~\cite{bodek2026jgres}. Further investigations are required to explain the specific physical mechanism that gives rise to this non-monotonic behavior in $\langle u_g \rangle^*$.

\subsection{Single Grain Properties Determine Granular Bed Strength }
\label{sec:singlegrain}

\begin{figure}[t!]
    \centering
    \includegraphics[width=\linewidth]{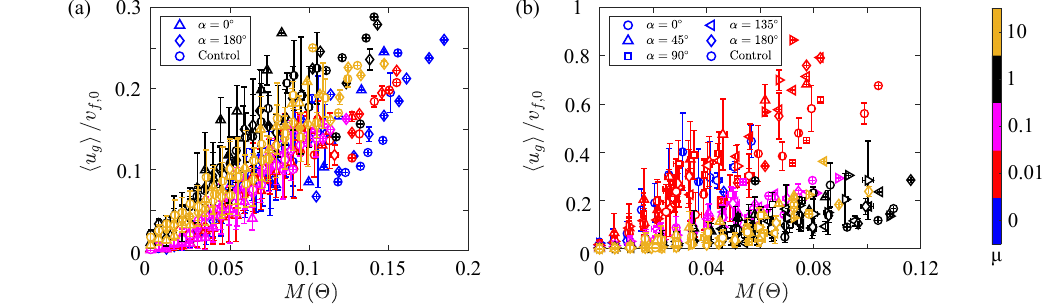}
    \caption{Normalized time-averaged mean surface grain velocity $\left\langle u_g \right\rangle / v_{f,0}$ plotted versus the fraction of mobile surface grains $M(\Theta)$ with $\Theta > \Theta_{g,c}$ for conditioning fluid flow directions $\alpha$ in (a) 2D and (b) 3D. In (a) and (b), the data are plotted for the conditioned bed at $\Theta_p / \Theta_c = 0.8$ and control bed for several static friction coefficients in the range $0 \le \mu \le 10$. Data are bin-averaged over $M(\Theta)$ with a bin width $0.005$ from $10$ randomly generated configurations for each $\mu$.}
    \label{fig:correlation}
\end{figure}

To understand the grain-scale mechanisms that give rise to anisotropic bed strength, we calculate the characteristic Shields number $\Theta_{g,c}$ above which individual grains at the bed surface become dislodged when all other grains are held fixed. We can then calculate the fraction of {\it mobile} surface grains $M(\Theta)$ that satisfy $\Theta \ge \Theta_{g,c}$. In Fig.~\ref{fig:correlation}, we show that $\left\langle u_g \right\rangle/v_{f,0}$ is approximately proportional to $M(\Theta)$ for all conditioning fluid flows (at angle $\alpha$) and static friction coefficients in 2D and 3D. The slope of $\left\langle u_g \right\rangle/v_{f,0}$ versus $M(\Theta)$ varies by $\sim 40\%$ over the range in static friction coefficients $0 \le \mu \le 10$ in 2D (Fig.~\ref{fig:correlation} (a)). The slope decreases by $\sim 90\%$ as $\mu$ increases from $0$ to $10$ in 3D (Fig.~\ref{fig:correlation} (b)). A lower slope indicates that grains entrained in the flow move more slowly. The large decrease in the slope of $\left\langle u_g \right\rangle/v_{f,0}$ versus $M(\Theta)$ in 3D compared to 2D is likely caused by the fact that a mobile surface grain in 3D can roll or slide around the grain in front of it, whereas a mobile surface grain in 2D must move over the grain in front of it.  Nonetheless, for all $\mu$, $\left\langle u_g \right\rangle$ is roughly linearly related to $M(\Theta)$ in both 2D and 3D, and thus, $M(\Theta)$ can be used to quantify the granular bed strength for all $\mu$.

\begin{figure}[t!]
    \centering
    \includegraphics[width=\linewidth]{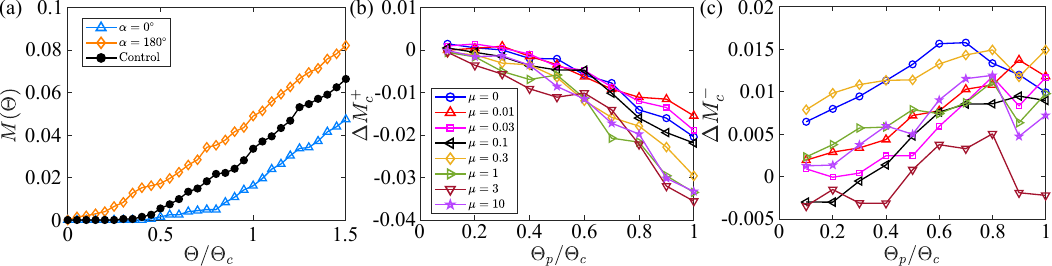}
    \caption{(a) The fraction of mobile surface grains $M(\Theta)$ plotted versus $\Theta/\Theta_c$ for a 2D conditioned bed at $\alpha = 0^{\circ}$ (blue triangles), $\alpha=180^{\circ}$ (orange diamonds), and the control bed (black solid circles). The granular bed is composed of grains with $\mu = 1$ and was conditioned at $\Theta_p / \Theta_c = 0.8$. (b) The difference $\Delta M_{c}^+=M_c(\alpha=0^{\circ})-M_c$ between $M_c$ for the conditioned bed at $\alpha = 0^{\circ}$ and the control bed plotted versus the Shields number of the preconditioning flow $\Theta_p / \Theta_c$ over a range of $\mu$. (c) The difference $\Delta M_{c}^-=M_c(\alpha=180^{\circ})-M_c$ between $M_c$ of the conditioned bed at $\alpha = 180^{\circ}$ and the control bed plotted versus $\Theta_p / \Theta_c$ for grains with the $\mu$ values in (b). Data in panels (b) and (c) are calculated as an ensemble average over $10$ randomly generated configurations for each $\mu$.}
    \label{fig:mobility_2d}
\end{figure}

We next investigate how the conditioning fluid flows affect $M(\Theta)$. In Fig.~\ref{fig:mobility_2d} (a), we show that $M(\Theta)$ for conditioned beds is larger (smaller) than that for the control bed for $\alpha = 180^{\circ}$ ($0^{\circ}$) over a wide range of $\Theta/\Theta_c$ after a granular bed with $\mu = 1$ has been conditioned with $\Theta_p / \Theta_c = 0.8$ in 2D. This result for $M(\Theta)$ highlights two important effects: (1) The conditioning flow dislodges grains that are unstable in the direction of the conditioning flow and brings these unstable grains to new locations on the bed surface where they can resist subsequent fluid flows. This process reduces the fraction of grains that can be subsequently dislodged in the direction of the conditioning flow and stabilizes the bed. (2) Many surface grains are unstable to fluid flows in the opposite direction of the conditioning fluid flow. These results emphasize that conditioned beds exhibit large bed strength anisotropy. To characterize the effects of the conditioning flow strength and inter-grain friction, we quantify the fraction of mobile grains at $\Theta = \Theta_c$ and compare that fraction for the conditioned and control beds: $\Delta M_c^+=M_c(\alpha=0^{\circ})-M_c$ and $\Delta M_c^-=M_c(\alpha=180^{\circ})-M_c$, where $M_c$ is the fraction of mobile grains at $\Theta=\Theta_c$ for the control bed. In Fig.~\ref{fig:mobility_2d} (b) and (c), we show that $\Delta M_c^+$ decreases and $\Delta M_c^-$ increases with increasing $\Theta_p$ for $\Theta_p/\Theta_c \lesssim 1$. The variations in $\Delta M_c^+$ and $\Delta M_c^-$ versus $\Theta_p$ highlight that conditioning flows cause significant changes to the structure of the bed surface and promote bed strength anisotropy. In addition, the dependence of $\Delta M_c^+$ and $\Delta M_c^-$ on $\Theta_p$ is similar for all $\mu$, which indicates that inter-grain friction does not strongly affect bed strength anisotropy in 2D. These results are qualitatively the same for system sizes $N \gtrsim 400$, as shown in Appendix~\ref{appendix:systemsize}.

\begin{figure}[t!]
    \centering
    \includegraphics[width=\linewidth]{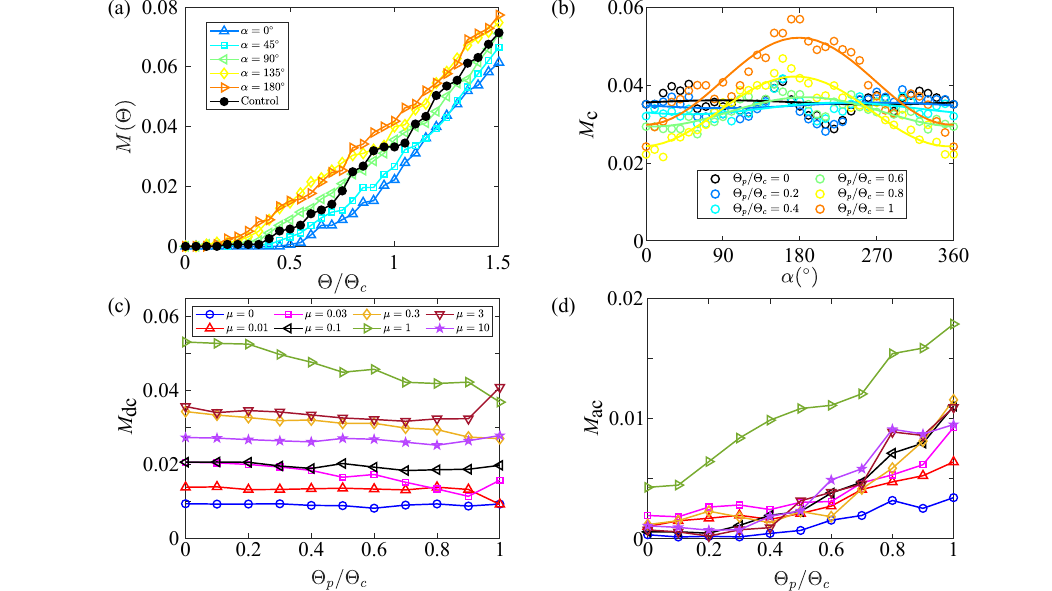}
    \caption{(a) The fraction $M(\Theta)$ of mobile surface grains with $\Theta \ge \Theta_{g,c}$ plotted versus $\Theta / \Theta_c$ for 3D conditioned beds at angle $\alpha$ and the control bed. The granular bed is composed of grains with inter-particle friction $\mu = 3$ and was conditioned by a fluid flow with Shields number $\Theta_p / \Theta_c = 0.8$. (b) The fraction of mobile surface grains $M_c$ at $\Theta =\Theta_c$ plotted versus $\alpha$ for a range of $\Theta_p/\Theta_c$ for the beds generated in (a). (c) The mean value $M_{dc}$ of $M_c(\alpha)$ and (d) amplitude $M_{ac}$ of the oscillations in $M_c(\alpha)$ plotted versus the Shields number of the preconditioning flow $\Theta_p / \Theta_c$ for different $\mu$. Data in panels (c) and (d) are calculated as ensemble average over $10$ randomly generated configurations for each $\mu$.}
    \label{fig:mobility_3d}
\end{figure}

In Fig.~\ref{fig:mobility_3d} (a), we show in 3D that $M(\Theta)$ of the conditioned bed is smaller (larger) at $\alpha = 0^{\circ}$ ($\alpha = 180^{\circ}$) than that for the control and $M(\Theta)$ increases monotonically with $0^{\circ} \le \alpha \le 180^{\circ}$. The variation of $M(\Theta)$ with $\alpha$ indicates that the number of grains that are susceptible to a fluid flow with a given $\Theta$ depends on $\alpha$. Unlike in 2D, $\alpha$ can be varied continuously in 3D. In Fig.~\ref{fig:mobility_3d} (b), we plot $M_c$ versus $\alpha$ for several values of $\Theta_p/\Theta_c$. We find that the amplitude of $M_c$ becomes larger with increasing $\Theta_p$ for $\Theta_p < \Theta_c$. Therefore, the bed strength anisotropy increases with $\Theta_p$ for $\Theta_p < \Theta_c$, similar to the results that we found in 2D.

To quantify the variation in $M_c$, we assume that $M_c(\alpha) = -M_{ac} \cos\alpha + M_{dc}$, such that the fraction of mobile grains is symmetric about the $x$-axis and $M_c$ has a period of $360^{\circ}$. $M_{dc}$ and $M_{ac}$ quantify the mean and standard deviation in $M_c$ and both $M_{dc}>0$ and $M_{ac}>0$ are positive. Fits of $M_c$ to the sinusoidal function are shown in Fig.~\ref{fig:mobility_3d} (b).  In Fig.~\ref{fig:mobility_3d} (c), we show that $M_{dc}$ is nearly constant with $\Theta_p$ for all $\mu$, except for $\mu = 1$, where $M_{dc}$ decreases by $\sim 20\%$ as $\Theta_p / \Theta_c$ increases from $0$ to $1$. However, we find that $M_{dc}$ at $\Theta_p / \Theta_c = 0$ varies with $\mu$, and we attribute this variation to the properties of the settled beds that are generated using the particular vibration procedure that we selected~\cite{silbert10softmatter}. The nearly constant behavior of $M_{dc}$ with $\Theta_p$ indicates that the conditioning flows do not alter the average bed strength, highlighting that compaction is minimal in the conditioned beds we study. In contrast, $M_{ac}$ increases with $\Theta_p$ for $0 \le \mu \le 10$, especially for $\Theta_p / \Theta_c \gtrsim 0.5$, as shown in Fig.~\ref{fig:mobility_3d} (d). $M_{ac}$ for $\mu = 1$ is noticeably larger than those at other $\mu$, which we attribute to the same cause of the decrease in $M_{dc}$ with $\Theta_p$. In Appendix.~\ref{appendix:vibration}, we show that increasing the vibrational frequency $\omega_b$ will reduce the variation of $M_{dc}$ and the overall magnitude of $M_{ac}$ as a function of $\Theta_p$ for $\mu = 1$. Nevertheless, $M_{ac}$ always increases with increasing $\Theta_p / \Theta_c$. Therefore, the increase in bed-strength anisotropy with $\Theta_p$ persists for all $\omega_b / \sqrt{K / m_s}$ studied, although the magnitude depends on $\omega_b / \sqrt{K / m_s}$. Unlike in 2D, we find that $M_{ac}$ is smaller for beds with $\mu = 0$ compared to those with $\mu > 0$, which suggests that inter-grain friction, though not necessary, can increase the bed strength anisotropy in 3D. These results are qualitatively the same for system sizes $N \gtrsim 400$, as shown in Appendix~\ref{appendix:systemsize}.

\subsection{Compaction in Granular Beds Caused by Conditioning Flows}
\label{sec:compaction}

\begin{figure}[t!]
    \centering
    \includegraphics[width=\linewidth]{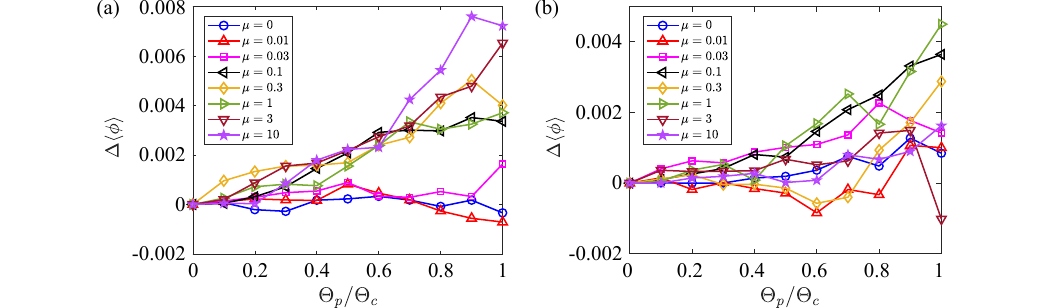}
    \caption{Change in the average local packing fraction $\Delta \left\langle \phi \right\rangle$ among surface grains in the initially settled bed and the bed following the conditioning fluid flow plotted versus the normalized conditioning Shields number $\Theta_p/\Theta_c$, for several inter-grain friction coefficients $\mu$ in (a) 2D and (b) 3D. The maximum compaction is $|\Delta \left\langle \phi \right\rangle| \lesssim 0.008$. Data are ensemble-averaged over $10$ randomly generated configurations for each $\mu$.}
    \label{fig:phi_change}
\end{figure}

Next, we investigate the role of compaction in the bed strength changes during conditioning flows. The protocol discussed in Sec.~\ref{sec:methods} generates densely packed settled granular beds, and thus we expect that the conditioning flows do not give rise to significant additional compaction. In Fig.~\ref{fig:phi_change}, we quantify the average change in the local packing fraction $\Delta \left\langle \phi \right\rangle$ of the surface grains between those in the control bed and those in the bed following the conditioning flows in 2D and 3D. We find that $\Delta \left\langle \phi \right\rangle \approx 0$ for granular beds with static friction coefficients $\mu \le 0.03$ for all $\Theta_p / \Theta_c$, while $|\Delta \left\langle \phi \right\rangle| \lesssim 0.008$ for $\mu > 0.03$ at $\Theta_p / \Theta_c \sim 1$ in 2D. In 3D, we find that $|\Delta \left\langle \phi \right\rangle| \lesssim 0.004$ at $\Theta_p / \Theta_c \sim 1$ for all $\mu$ studied. We attribute the small variations in $\Delta \left\langle \phi \right\rangle$ at $\Theta_p / \Theta_c \sim 1$ for different $\mu$ to the slightly different degrees of compaction induced by this particular vibration protocol~\cite{silbert10softmatter}. We note that $|\Delta \left\langle \phi \right\rangle|$ in the current studies is much smaller than $|\Delta \left\langle \phi \right\rangle| \sim 0.06$ for $\Theta_p / \Theta_c \sim 1$ obtained from other studies of sedimented beds where compaction played a significant role in determining bed strength~\cite{allen2018prf}. In Appendix~\ref{appendix:compaction}, we show that $\Delta\left\langle\phi\right\rangle$ caused by the conditioning flows is much smaller for control beds that were vibrated compared to that for control beds that were not vibrated. Therefore, compaction caused by the conditioning flows is minimal and we will show below that changes in the inter-grain fabric give rise to anisotropic bed strength.  

\subsection{Grain-scale Fabric Determines Strength of Bed Surface}
\label{sec:fabric}

After showing that conditioning fluid flows promote anisotropy in the granular bed strength, i.e., $\Theta_{g,c}$ and $F_{g,c}$ vary with the conditioning flow direction $\alpha$, we next identify the changes in the structural and mechanical properties of granular beds that cause $\Theta_{g,c}$ and $F_{g,c}$ to vary with the direction of the fluid flow. Here, we show that the critical force to dislodge surface grain $i$ can be approximated using an analytical expression in both 2D and 3D if the positions of grain $i$ and its neighbors are known. We can then validate the analytical expression for the minimum force $F_{g,c}^0$ to dislodge a grain with the value $F_{g,c}$ obtained from DEM simulations that determine the smallest force necessary to move grain $i$ by more than $\sigma_l$.

\begin{figure}[t!]
    \centering
    \includegraphics[width=\linewidth]{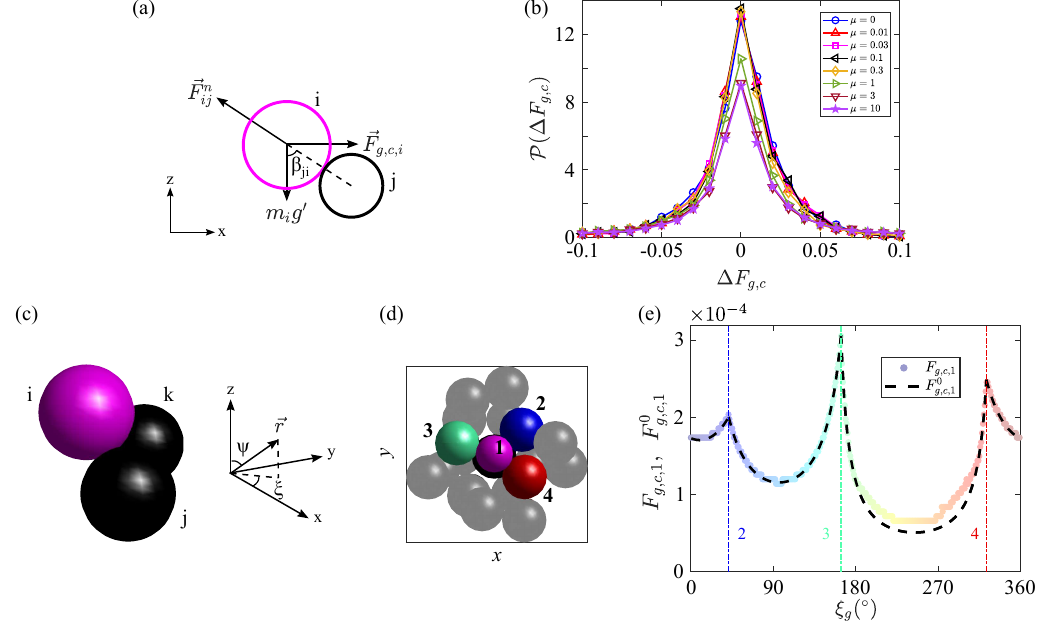}
    \caption{(a) Illustration of the calculation of the critical force $F_{g,c,i}$ to dislodge grain $i$ (magenta circle) that is in contact with grain $j$ (black circle) with the associated Cartesian coordinate axes in 2D. $\vec{F}_{ij}^n$ is the normal component of the force on grain $i$ from grain $j$, $m_ig'$ is the magnitude of the gravitational force on grain $i$, and $\beta_{ji}$ is the angle between the unit vector $\hat{r}_{ji}$ pointing from the center of grain $i$ to the center of grain $j$ and $-{\hat z}$. (b) Probability distribution $\mathcal{P}(\Delta F_{g,c})$ of the deviation $\Delta F_{g,c} = F_{g,c} / F_{g,c}^0 - 1$, where $F_{g,c}$ is the minimum force required to dislodge a grain from the DEM simulations and $F_{g,c}^0$ is calculated analytically in  Eq.~\ref{eq:fc_2d} for 2D granular beds with inter-grain static friction coefficient, $\mu$, indicated by the colors and symbols. The data are obtained from $10$ independent configurations, each conditioned at Shields number $0.1 \le \Theta_p / \Theta_c \le 1$. (c) Illustration of the calculation of the critical force $F_{g,c,i}$ to move a surface grain in 3D. (Left panel) An example bed surface, consisting of the test grain $i$ (magenta sphere) and two contacting grains $j$ and $k$ (black spheres). (Right panel) Illustration of the polar $\Psi$ and azimuthal $\xi$ angles of position vector $\vec{r}$ in spherical coordinates. (d) A small region of the surface of a granular bed consisting of a central surface grain $1$ (magenta sphere) that is in contact with three other grains $2$, $3$, and $4$ (blue, green, and red spheres), and has several next nearest neighbors (gray spheres). (e) Minimum force required to dislodge grain $1$ from the granular bed shown in panel (d), calculated analytically from Eq.~\ref{eq:fc_3d} $F_{g,c,1}^0$ (black dashed line) and obtained from DEM simulations $F_{g,c,1}$ (filled circles), plotted versus the azimuthal angle of the force $\xi_{g}$. The colors for grains $2$, $3$, and $4$ in (d) indicate the azimuthal angles of $\hat{r}_{21}$, $\hat{r}_{31}$, and $\hat{r}_{41}$ with the same color scheme as $F_{g,c,1}(\xi_g)$ in (e). The vertical dash-dotted lines in (e) give the azimuthal angles of $\hat{r}_{21}$, $\hat{r}_{31}$, and $\hat{r}_{41}$.}
    \label{fig:contact_angle}
\end{figure}

Assuming frictionless granular beds in 2D (i.e. zero tangential forces), for test grain $i$ in contact with grain $j$ and assuming grain $j$ is fixed, we show in Appendix~\ref{appendix:fc_formula} that the critical force to dislodge grain $i$ can be expressed as 
\begin{equation}
    F_{g,c,i}^0 = m_i g' \tan\beta_{ji},
\label{eq:fc_2d}
\end{equation}
where $\beta_{ji}$ is the angle between the unit vector $\hat{r}_{ji}$ pointing from the center of grain $i$ to the center of grain $j$ and $-{\hat z}$ as shown in Fig.~\ref{fig:contact_angle} (a). In Fig.~\ref{fig:contact_angle} (b), we find that $F_{g,c}^0$ matches $F_{g,c}$ to within $5\%$ for $0 \le \mu \le 10$, even though $F_{g,c}^0$ was obtained for $\mu = 0$. We attribute the small discrepancy between the results for the MD simulations and the analytical expression to fact that next nearest neighbor grains can potentially prevent the test grain from moving a distance $\sigma_l$.  

In 3D, unlike in 2D, if the test grain $i$ is only in contact with grain $j$, then any applied force ${\vec F}_{g,c,i}$ in the $x$-$y$ plane will not be force-balanced unless $\vec{F}_{g,c,i}$ is in the plane formed by $\hat{r}_{ji}$ and $-{\hat z}$. Therefore, to calculate ${\vec F}_{g,c,i}$, grain $i$ needs to be in contact with at least $2$ other grains, e.g. grains $j$ and $k$ in Fig.~\ref{fig:contact_angle} (c), and we need to specify the polar $\Psi$ and azimuthal $\xi$ angles for the position vector $\vec{r}$, as shown in Fig.~\ref{fig:contact_angle} (c). Under the assumption that all grains in contact with grain $i$ are fixed and that the grains are frictionless, we derive the following expression for $F_{g,c,i}^0$ in Appendix~\ref{appendix:fc_formula}:
\begin{equation}
\label{eq:fc_3d}
    F_{g,c,i}^0 = m_i g' \frac{\cos(\Delta\xi_{ki,kj}) \sin(\Delta\xi_{g,kj} - \Delta\xi_{ij,kj}) - \cos(\Delta\xi_{ij,kj}) \sin(\Delta\xi_{g,kj} - \Delta\xi_{ik,kj})}{\sin\Psi_{ik} \cos\Psi_{ij} \sin(\Delta\xi_{g,kj} - \Delta\xi_{ik,kj}) - \sin\Psi_{ij} \cos\Psi_{ik} \sin(\Delta\xi_{g,kj} - \Delta\xi_{ij,kj})} \frac{\sin\Psi_{ij} \sin\Psi_{ik}}{\cos(\Delta\xi_{g,kj})},
\end{equation}
where $\xi_g$ is the azimuthal angle of $\vec{F}_{g,c,i}^0$, $\Psi_{ij}$ and $\xi_{ij}$ are the polar and azimuthal angles of $\hat{r}_{ij}$, $\Delta\xi_{g, kj} = \xi_g - \xi_{kj} + \pi / 2$, $\Delta\xi_{ij, kj} = \xi_{ij} - \xi_{kj} + \pi / 2$, and $\Delta\xi_{ik, kj} = \xi_{ik} - \xi_{kj} + \pi / 2$. We study an example region of a granular bed surface with $\mu = 1$ in Fig.~\ref{fig:contact_angle} (d), where grain $1$ has $3$ contacts (grains $2$, $3$, and $4$). We compare Eq.~\ref{eq:fc_3d} for $F_{g,c,1}^0$ to $F_{g,c,1}$ obtained from DEM simulations in Fig.~\ref{fig:contact_angle} (e). $F_{g,c,1}^0$ for $\xi_g$ in the range between the vertical dashed lines marked by the numbers $2$ and $3$ is calculated based on the positions of grains $1$, $2$, and $3$. Similarly, $\xi_g$  between the vertical dashed lines marked by the numbers $3$ and $4$ indicate that $F_{g,c,1}^0$ was calculated from the positions of grains $1$, $3$, and $4$. We observe an excellent match between $F_{g,c,1}^0$ and $F_{g,c,1}$ for all angles $\xi_g$. Nearest neighbor grains may also prevent test grain $i$ from moving a distance $\sigma_l$ and cause small deviations between $F_{g,c,i}^0$ and $F_{g,c,i}$ in 3D.

We find excellent agreement between $F_{g,c}$ from the MD simulations over a wide range of $\mu$ and the analytical expressions in Eqs.~\ref{eq:fc_2d} and~\ref{eq:fc_3d} in 2D and 3D. However, in the MD simulations, the normal and tangential forces on the surface grains are nonzero and vary over the inter-grain contacts.  We postulate that the tangential forces do not play a significant role in determining $F_{g,c}$ because the contacting grains can roll over neighboring grains instead of slipping. Also, we hypothesize that the normal forces on the surface grains do not play a significant role in determining $F_{g,c}$ because the target grain only makes one or  two contacts when it is on the verge of being dislodged in 2D and 3D. Therefore, these results suggest that the grain-scale fabric, not the force network controls the incipient strength of the bed surface.  In contrast, grain-scale stresses can play a more important role setting the grain speed after it is dislodged from the bed surface and initiating the dynamics of larger regions of the bed. Finally, we note that the calculation of $F_{g,c}$ to dislodge a single grain is obtained under the assumption that all other grains are held fixed. We believe that this assumption is reasonable since we focus on the regime $\Theta \lesssim \Theta_c$ where grains independently become dislodged, instead of becoming dislodged simultaneously as a group of grains. Collective grain motion that leads to grains becoming entrained in the flow will be considered in future studies to fully understand the dynamics of granular beds driven by fluid flows.

\section{Conclusions and Future Directions}
\label{sec:summary}

Previous experimental studies have shown that conditioning a granular bed using sub-critical fluid flows can alter the strength of the granular bed. Specifically, the conditioned bed becomes stronger with respect to fluid flows in the same direction of the conditioning flow, but weaker in response to fluid flows in the opposite direction. In this work, we employ DEM simulations of fluid-driven granular beds to elucidate the grain-scale changes in the bed that give rise to anisotropic bed strength in 2D and 3D.

First, we show that the mean surface grain speed $\left\langle u_g \right\rangle$ of conditioned beds is smaller (larger) in the same (opposite) direction of the conditioning flows compared to that of the control bed over a wide range of inter-grain static friction coefficients ($0 \le \mu \le 10$) in both 2D and 3D. This result in 2D shows that conditioning fluid flows can induce directional strengthening and weakening even in extremely narrow beds. In the current work, we focus on already compacted beds, and thus conditioning fluid flows do not lead to additional compaction.  Instead, we find that the fraction of mobile surface grain $M(\Theta)$ is highly correlated with $\left\langle u_g \right\rangle$, which suggests that the critical Shields number $\Theta_{g,c}$ to dislodge a surface grain, can be used to explain the bed strength anisotropy. In 2D, $M(\Theta)$ is lower (higher) at $\alpha = 0^{\circ}$ ($\alpha = 180^{\circ}$), and the difference in $M(\Theta)$ at $\alpha = 0^{\circ}$ and $180^{\circ}$ increases with the Shields number of the conditioning flow for $0.5 \lesssim \Theta_p \lesssim \Theta_c$ independent of $\mu$. In 3D, $M(\Theta)$ can be described by a sinusoidal function with a period of $360^{\circ}$, nearly constant mean $M_{dc}$, and oscillation amplitude $M_{ac}$ that increases with $\Theta_p$. $M_{ac}$ is non-zero for $\mu = 0$ and grows for $\mu > 0$ in 3D, which indicates that inter-grain static friction promotes bed strength anisotropy. Finally, we show that the minimum force $F_{g,c}$ to dislodge a surface grain when holding all other grains fixed can be calculated using only the relative orientations of the contacts, while inter-grain normal and tangential forces do not play significant roles in determining $F_{g,c}$. Further, the DEM simulations in the present work allow researchers to probe the grain-scale mechanisms that determine the strength of granular beds. 

These findings raise several interesting directions for future research. First, we showed that the minimum force required to dislodge a surface  grain in a granular bed is determined by the inter-grain contact network when holding all other grains fixed. When do inter-grain forces, including both their magnitudes and directions, matter? One possible scenario is that inter-grain forces play a more important role when grains are entrained in the flow (for $\Theta \sim \Theta_c$) and collide with static surface grains giving rise to momentum transfer. Another possible scenario is that grains with larger $F_{g,c}$ can transmit fluid-born forces to grains with smaller $F_{g,c}$, causing collective rearrangements so that multiple grains can become dislodged at the same time. Second, in many applications and in natural contexts, granular materials are composed of non-spherical grains. We can extend the calculations of the minimal force required to dislodge a grain to non-spherical grains, which likely requires testing all possible relative orientations between the grain and its contacting grains since non-spherical grains can rotate away from their initial contacts to more or less stable configurations. Also, in the current study, we did not consider turbulent fluid flows, which include flow speed fluctuations over a wide range of time and length scales. Interesting questions for turbulent fluid flows include: (1) do fluctuations from turbulent conditioning flows promote or suppress bed strength anisotropy compared to viscous conditioning flows and (2) how does the energy-wavenumber cascade in turbulent flows affect the correlations between structure and anisotropic bed strength? Finally, non-horizontal granular beds are common in nature, and the relationship $M_c(\alpha) = -M_{ac} \cos \alpha + M_{dc}$ will change for inclined beds. In particular, both $M_{dc}$ and $M_{ac}$ can vary with the angle of inclination of the bed.

\begin{acknowledgments}

We thank Dr. Abe Clark for helpful discussions. We acknowledge support from the Army Research Office under Grant No. W911NF-23-1-0032 and No. W911NF-25-1-0158, as well as from NSF PHY-2309135 to the Kavli Institute for Theoretical Physics (KITP). This work was also supported by the High Performance Computing facilities operated by Yale’s Center for Research Computing.

\end{acknowledgments}

\section*{Data Availability}

The data that support the findings of this article are publicly available~\cite{wang2026data}.

\appendix

\section{Calculation of the Critical Shields Number $\Theta_c$}
\label{appendix:theta_c}

\begin{figure}[h!]
    \centering
    \includegraphics[width=\linewidth]{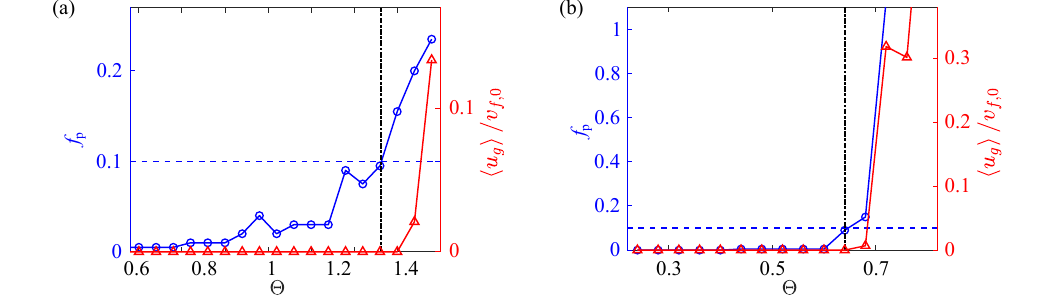}
    \caption{Fraction $f_p$ of surface grains that have moved more than $\sigma_l$ after applying a fluid flow (left, blue circles), and normalized mean surface grain speed $\left\langle u_g \right\rangle / v_{f,0}$ averaged over the final $1\%$ of the total duration of the flow (right, red triangles), plotted versus the Shields number $\Theta$ of the fluid flow in (a) 2D and (b) 3D. The inter-grain static friction coefficient $\mu = 1$ for the granular beds, the horizontal blue dashed line corresponds to $f_p = 0.1$, and the vertical black dashed-dotted line corresponds to $\Theta = \Theta_c$ in (a) and (b).}
    \label{fig:theta_c}
\end{figure}

In this Appendix, we describe the calculation of the critical Shields number $\Theta_c$ for the control beds. As discussed in Sec.~\ref{sec:methods}, we set $\Theta = \Theta_c$ when the following two criteria are satisfied: (1) the four stopping criteria in Sec.~\ref{sec:methods} for each of the grains are satisfied and (2) less than $10\%$ of the surface grains have moved more than $\sigma_l$. We show the fraction $f_p$ of surface grains that have moved more than $\sigma_l$ and the mean surface grain speed $\left\langle u_g \right\rangle$ (averaged over the final $1\%$ of the total duration of the fluid flow) in Fig.~\ref{fig:theta_c}. We define $\Theta_c$ to be the largest $\Theta$ where $f_p \le 10\%$ and $\left\langle u_g \right\rangle = 0$, as indicated by vertical black dashed-dotted lines in Fig.~\ref{fig:theta_c}. We note that $f_e$ continues to increase as $\Theta > \Theta_c$, indicating that the bed has undergone grain rearrangements. $\Theta_c$ can increase for granular beds with significant rearrangements, leading to decreasing $\left\langle u_g \right\rangle / v_{f,0}$ with increasing $\Theta$ (not shown in Fig.~\ref{fig:theta_c}). However, we are interested in the smallest $\Theta_c$ as we increase $\Theta$ from zero since we compare the results for control beds to those for conditioned beds with $\Theta_p / \Theta_c \le 1$.

\section{Time Dependence of Fluid-driven Grain Dynamics}
\label{appendix:timescale}

\begin{figure}[t!]
    \centering
    \includegraphics[width=\linewidth]{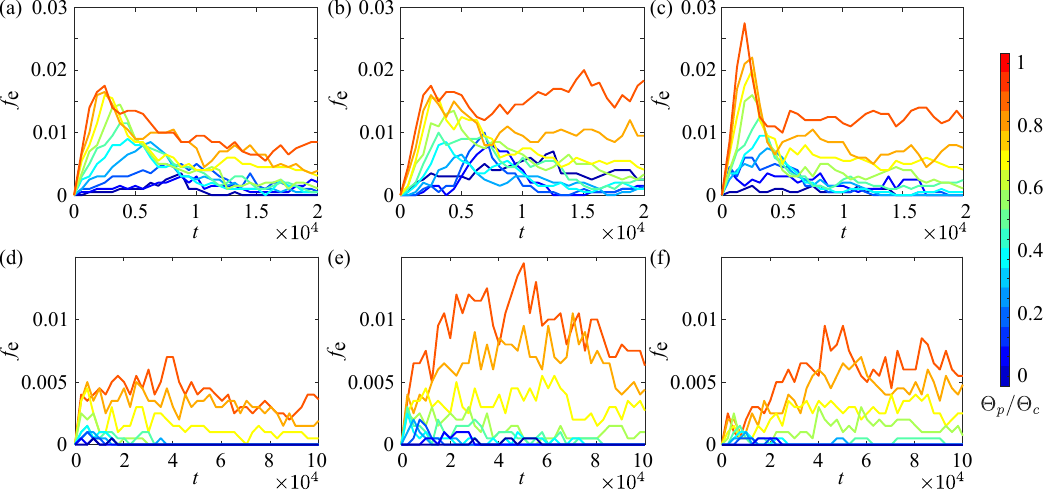}
    \caption{Fraction $f_e$ of surface grains that have been entrained in the fluid flow (i.e. grains with speeds greater than $v_{f,0}$) plotted versus the time $t$ that the conditioning fluid flow was applied in (a)-(c) 2D and (d)-(f) 3D. The inter-grain static friction coefficient $\mu=0$ in (a) and (d), $0.3$ in (b) and (e), and $3$ in (c) and (f). The Shields number $\Theta_p / \Theta_c$ of the conditioning flows are indicated by the colors increasing from blue to red.}
    \label{fig:mobile_precond}
\end{figure}

To investigate how the grain dynamics evolve as a function of time, we focus on surface grains that have been entrained in the fluid flow, i.e. grains that have speeds greater than $v_{f,0}$ in the flow direction. In Fig.~\ref{fig:mobile_precond}, we show the time dependence of the fraction of entrained grains $f_e$ for granular beds with a range of $\mu$ in 2D and 3D. In general, $f_e$ first increases and then decreases to $0$ with time $t$, for all $\mu$ and $\Theta_p / \Theta_c$. The total time for this initial increase and subsequent decrease such that $f_e \rightarrow 0$ is approximately $T_w \sim 10^4$ in 2D and $5 \times 10^4$ in 3D at low $\Theta_p/\Theta_c \lesssim 0.7$. For $\Theta_p / \Theta_c \gtrsim 0.7$, $f_e$ has a rapid increase and then decreases to a quasi-plateau value for $t<T_w$. The quasi-plateau decreases more slowly to zero over a timescale of $\sim 10T_w$. Therefore, we select $T_w^* = 10^4$ in 2D and $T_w^* = 5 \times 10^4$ in 3D as the averaging time window size for calculating $\langle u_g \rangle$ and other quantities that characterize bed strength anisotropy.

\section{Compaction in Vibrated versus Unvibrated Granular Beds}
\label{appendix:compaction}

\begin{figure}[t!]
    \centering
    \includegraphics[width=\linewidth]{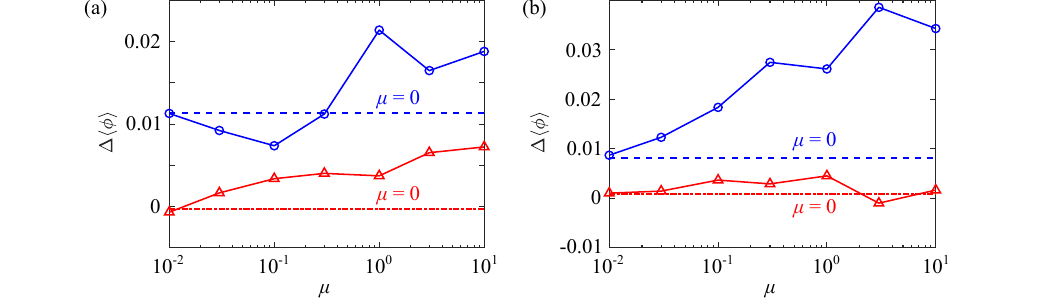}
    \caption{Change in the average local packing fraction $\Delta \left\langle \phi \right\rangle$ among surface grains in the settled bed without adding vibrations and those in the settled bed with vibrations (blue symbols) and between surface grains in the initially settled bed and those following the conditioning fluid flow at normalized conditioning Shields number $\Theta_p/\Theta_c = 1$ (red symbols) plotted versus the inter-grain static friction coefficient $\mu$ in (a) 2D and (b) 3D. The dashed and dashed-dotted horizontal lines represent $\Delta \left\langle \phi \right\rangle$ at $\mu = 0$ in (a) and (b).  Data are ensemble-averaged over $10$ randomly generated configurations for each $\mu$.}
    \label{fig:phi_compaction}
\end{figure}

In Fig.~\ref{fig:phi_change} in the main text, where vibration was used to generate the settled granular beds, we showed that the change in the average packing fraction from the conditioning flows is small $\Delta\left\langle\phi\right\rangle < 0.008$. To further emphasize that the compaction caused by the conditioning flows does not affect bed strength anisotropy, we generated another set of settled beds without using the vibration protocol. In particular, we compare the mean packing fraction difference $\Delta\left\langle\phi\right\rangle$ of surface grains in beds generated with and without the vibration protocol. In Fig.~\ref{fig:phi_compaction}, we show that $\Delta\left\langle\phi\right\rangle$ (for vibrated versus unvibrated beds) is much larger compared to that induced by the conditioning flows at $\Theta_p / \Theta_c = 1$ over the full range $0 \le \mu \le 10$. In particular, $\Delta\left\langle\phi\right\rangle$ caused by vibration is approximately twice that caused by the conditioning flows in 2D and more than five times that caused by the conditioning flows in 3D. Therefore, compaction induced by the conditioning flows is minimal.

\section{Effect of the Vibration Frequency on the Control Bed Strength}
\label{appendix:vibration}

\begin{figure}[h!]
    \centering
    \includegraphics[width=\linewidth]{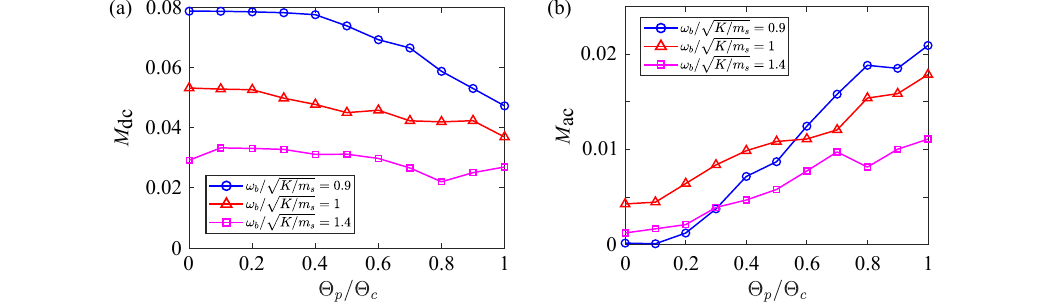}
    \caption{(a) The mean value $M_{dc}$ and (b) amplitude $M_{ac}$ of the oscillations in $M_c(\alpha)$ plotted versus the Shields number of the preconditioning fluid flow $\Theta_p / \Theta_c$ for granular beds generated using vibration at fixed amplitude $z_0=\sigma_s$ and several vibrational frequencies $\omega_b$. The inter-grain static friction coefficient is $\mu = 1$. Data are calculated as an ensemble average over $10$ randomly generated granular beds for each $\omega_b$.}
    \label{fig:vibration}
\end{figure}

In Fig.~\ref{fig:mobility_3d} in the main text, we showed that the average $M_{dc}(\Theta_p)$ and fluctuations $M_{ac}(\Theta_p)$ in the fraction of mobile grains possessed nonmonotonic dependence on the static friction coefficient $\mu$ for 3D granular beds. In particular, granular beds with $\mu = 1$ had larger values for $M_{dc}$ and $M_{ac}$ compared to those for all other $\mu$. We hypothesize that the vibrational frequency $\omega_b$ used to generate static beds impacts the structure of the sedimented bed differently for each $\mu$~\cite{an2008powdertech, gago2016paperphys}. To test this hypothesis, we studied two additional values of $\omega_b$ (at fixed amplitude $z_0=\sigma_s$) for granular beds with $\mu = 1$: $\omega_b/\sqrt{K/m_s} = 0.9$ and $1.4$, in addition to $1$ studied in the main text. We find that $M_{dc}$ at $\Theta_p = 0$ decreases with increasing $\omega_b$, which indicates that the control bed becomes stronger with increasing vibration frequencies (Fig.~\ref{fig:vibration} (a)). Furthermore, we find that the magnitude of the decrease in $M_{dc}(\Theta_p)$ from $\Theta_p / \Theta_c = 1$ to $0$ deceases with increasing $\omega_b$, and that $M_{dc}(\Theta_p)$ remains constant for $0 \le \Theta_p / \Theta_c \le 1$ for $\omega_b / \sqrt{K / m_s} = 1.4$. Similarly, we show that the magnitude of the increase in $M_{ac}(\Theta_p)$ from $\Theta_p / \Theta_c = 0$ to $1$ decreases with increasing $\omega_b$ (Fig.~\ref{fig:vibration} (b)). Together, these results suggest that the changes in $M_{dc}(\Theta_p)$ and $M_{ac}(\Theta_p)$ depend on $\mu$ {\it and} $\omega_b$. Note that the behavior of $M_{dc}(\Theta_p)$ and $M_{ac}(\Theta_p)$ at $\omega_b / \sqrt{K / m_s} = 1.4$ is similar to results shown in Figs.~\ref{fig:mobility_3d} (c) and (d) for $\mu$ other than $1$ at $\omega_b / \sqrt{K / m_s} = 1$. Nevertheless, the increase of $M_{ac}$ with increasing $\Theta_p$ for all studied $\omega_b$ shows that the conditioning fluid flows give rise to bed strength anisotropy. Therefore, the increase in bed-strength anisotropy with $\Theta_p$ persists for all $\omega_b / \sqrt{K / m_s}$ studied, although the magnitude depends on $\omega_b / \sqrt{K / m_s}$.

\section{System-Size Dependence in 2D and 3D}
\label{appendix:systemsize}

\begin{figure}[t!]
    \centering
    \includegraphics[width=0.9\linewidth]{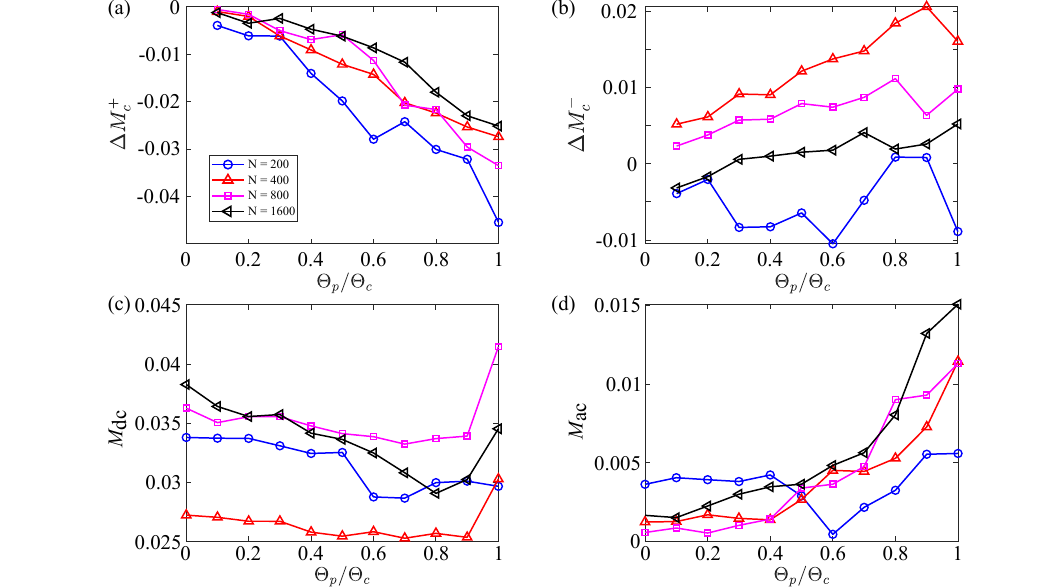}
    \caption{(a) The difference $\Delta M_{c}^+=M_c(\alpha=0^{\circ})-M_c$ and (b) $\Delta M_{c}^-=M_c(\alpha=180^{\circ})-M_c$ between the fraction of mobile surface grains $M_c$ for the control bed and the conditioned bed at (a) $\alpha = 0^{\circ}$ and (b) $\alpha = 180^{\circ}$ plotted versus the Shields number of the preconditioning flow $\Theta_p / \Theta_c$ over a range of system sizes $N$ with inter-grain friction coefficient $\mu = 1$ in 2D. (c) The mean value $M_{dc}$ of $M_c(\alpha)$ and (d) amplitude $M_{ac}$ of the oscillations in $M_c(\alpha)$ plotted versus $\Theta_p / \Theta_c$ for different $N$ with inter-grain friction coefficient $\mu = 3$ in 3D. The data in all panels are calculated as an ensemble average over 10 independently generated beds for each $N$.}
    \label{fig:systemsize}
\end{figure}

In this Appendix, we study the system-size effects in both 2D and 3D with $N = 200$, $400$, $800$ and $1600$. We show results for the fraction of mobile grains at the critical Shields number as a function of system size in Fig.~\ref{fig:systemsize}. In 2D, we find that $\Delta M_{c}^+$ is negative for all $\Theta_p / \Theta_c$ and $\Delta M_{c}^+$ decreases with increasing $\Theta_p / \Theta_c$ for all $N$ studied (Fig.~\ref{fig:systemsize} (a)), which emphasizes that the conditioning flows strengthen the granular bed. However, $\Delta M_{c}^-$ remains approximately zero over the full range of $\Theta_p / \Theta_c$ for $N = 200$ and only increases with increasing $\Theta_p / \Theta_c$ for $N \ge 400$ (Fig.~\ref{fig:systemsize} (b)). In 3D, $M_{dc}$ remains nearly constant over the full range of $\Theta_p / \Theta_c$ for all $N$ (Fig.~\ref{fig:systemsize} (c)). Further, $M_{ac}$ does not change with $\Theta_p / \Theta_c$ for $N = 200$ and only increases with increasing $\Theta_p / \Theta_c$ for $N \ge 400$ (Fig.~\ref{fig:systemsize} (d)). Here, we show that the results for the fraction of mobile grains are qualitatively the same for $N \ge 400$ in both 2D and 3D.

\section{Derivation of Minimum Force to Dislodge a Frictionless Grain in 2D and 3D}
\label{appendix:fc_formula}

In this Appendix, we provide additional details for the derivation of the minimum force $F_{g,c}^0$ to move a grain in frictionless granular beds in both 2D and 3D. Similar analysis techniques have been used to predict the critical Shields number for a sphere pinned against a U-shaped rod submerged in fluids as a function of their viscosity~\cite{kudrolli2016jfm}.

In 2D, for test grain $i$ in contact with grain $j$, there are three forces acting on $i$: the applied force $F_{g,c,i}$ to move the grain, the normal repulsive force $F_{ij}^n$ from grain $j$, and the gravitational force $m_i g'$, as illustrated in Fig.~\ref{fig:contact_angle} (a). For $F_{g,c,i}$ to move grain $i$, the following two force-balance equations in the $x$- and $z$-directions must be satisfied:
\begin{equation}
    \label{eq:fc_2d_1}
    F_{g,c,i} \ge F_{ij}^n \sin\beta_{ji},
\end{equation}
and
\begin{equation}
    \label{eq:fc_2d_2}
    F_{ij}^n \cos\beta_{ji} \ge m_i g',
\end{equation}
where $\beta_{ji}$ is the angle between the unit vector $\hat{r}_{ji}$ from the center of grain $i$ to the center of grain $j$ and $-{\hat z}$ and $0 \le \beta_{ji} < \pi/2$, as illustrated in Fig.~\ref{fig:contact_angle} (a). Eqs.~\ref{eq:fc_2d_1} and~\ref{eq:fc_2d_2} enforce that
\begin{equation}
    \label{eq:fc_2d_3}
    F_{g,c,i} \ge F_{g,c,i}^0 = m_i g' \tan\beta_{ji}.
\end{equation}
It is immediately clear that $\beta_{ji}$ becomes smaller once grain $i$ starts moving (to the right in Fig.~\ref{fig:contact_angle} (a)). No solution exists for $F_{g,c,i}^0$ when $\beta_{ji} \ge \pi / 2$. That is, grain $i$ cannot be moved regardless of the magnitude of the applied force when $\beta_{ji} \ge \pi / 2$ with grain $j$ fixed.

In 3D, we consider the test grain $i$ to be in contact with two grains $j$ and $k$ (Fig.~\ref{fig:contact_angle} (c)). In this scenario, there are four forces acting on $i$: the applied force $F_{g,c,i}$ to move grain $i$, the normal repulsive forces $F_{ij}^n$ and $F_{ik}^n$ from grains $j$ and $k$, respectively, and the gravitational force $m_i g'$. We ensure that grains $j$ and $k$ are ordered counter-clockwise with respect to grain $i$ when their center positions are projected onto the $x$-$y$ plane. For any vector $\vec{r}$, we denote its associated polar and azimuthal angles in spherical coordinate as $\Psi$ and $\xi$, respectively, shown in Fig.~\ref{fig:contact_angle} (c). For $F_{g,c,i}$ with azimuthal angle $\xi_g$ to move grain $i$, the following three equations representing force balance in the $z$-direction and directions parallel and perpendicular to $\vec{r}_{kj, xy} = \vec{r}_{kj} - (\vec{r}_{kj} \cdot \hat{z}) \hat{z}$ have to be satisfied:
\begin{equation}
    \label{eq:fc_3d_1}
    F_{ij} \cos\Psi_{ij} + F_{ik} \cos\Psi_{ik} \ge m_i g',
\end{equation}
\begin{equation}
    \label{eq:fc_3d_2}
     F_{g,c,i} \sin(\Delta\xi_{g, kj}) + F_{ij} \sin\Psi_{ij} \sin(\Delta\xi_{ij, kj}) + F_{ik} \sin\Psi_{ik} \sin(\Delta\xi_{ik, kj}) = 0,
\end{equation}
and 
\begin{equation}
    \label{eq:fc_3d_3}
    F_{g,c,i} \cos(\Delta\xi_{g,kj}) \ge F_{ij} \sin\Psi_{ij} \cos(\Delta\xi_{ij, kj}) + F_{ik} \sin\Psi_{ik} \cos(\Delta\xi_{ik, kj}).
\end{equation}
The azimuthal angle must satisfy $\xi_{ij} \le \xi_g \le \xi_{ik}$; otherwise, grain $i$ will lose contact with grain $j$ ($\xi_g > \xi_{ik}$) or $k$ ($\xi_g < \xi_{ij}$). Equations~\ref{eq:fc_3d_1} - \ref{eq:fc_3d_3} enforce that
\begin{equation}
    \label{eq:fc_3d_4}
    \begin{split}
    F_{g,c,i} & \ge F_{g,c,i}^0 \\
    & = m_i g' \frac{\cos(\Delta\xi_{ki,kj}) \sin(\Delta\xi_{g,kj} - \Delta\xi_{ij,kj}) - \cos(\Delta\xi_{ij,kj}) \sin(\Delta\xi_{g,kj} - \Delta\xi_{ik,kj})}{\sin\Psi_{ik} \cos\Psi_{ij} \sin(\Delta\xi_{g,kj} - \Delta\xi_{ik,kj}) - \sin\Psi_{ij} \cos\Psi_{ik} \sin(\Delta\xi_{g,kj} - \Delta\xi_{ij,kj})} \frac{\sin\Psi_{ij} \sin\Psi_{ik}}{\cos(\Delta\xi_{g,kj})}.
    \end{split}
\end{equation}
There is no solution for $F_{g,c,i}^0$ when both $\Psi_{ij} \ge \pi / 2$ and $\Psi_{ik} \ge \pi / 2$. That is, grain $i$ cannot be moved regardless of the magnitude of the applied force when $\Psi_{ij} \ge \pi / 2$ and $\Psi_{ik} \ge \pi / 2$ with grains $j$ and $k$ fixed.

\bibliography{biblio}

\end{document}